\documentclass{ceab}  

\usepackage{epsfig}    
\usepackage{graphicx}   

\usepackage{ceabbib}     

\usepackage[T1]{fontenc}

\begin{document}

\title{Considerations on the Sp\"orer-diagram - torsional wave relationship}

\author{J. Murak\"ozy and A. Ludm\'any
\vspace{2mm}\\
\it Heliophysical Observatory, Hungarian Academy of Sciences, \\
\it H-4010 Debrecen, P.O.Box 30. Hungary}

\maketitle

\begin{abstract}

Studies on the earlier reported spatial correlations between magnetic field distributions and the torsional wave have been substantially extended to the period 1975-2009. The investigations are based on Debrecen sunspot data and magnetic field data of Mount Wilson. The recently available data and distributions seem to support the idea that the torsional wave may be resulted in a flow round the toruses which, in turn, results in Coriolis-deviations forward and backward, establishing the prograde and retrograde belts. 

\end{abstract}

\keywords{sunspots, torsional wave}

\def\gore{torsional wave - sunspots}

\section{Introduction}

Ever since the first results about the torsional pattern reported by Howard and LaBonte (1980) the interpretation of this unexpected phenomenon has been a challenge. The similarity of the equatorward migration of these belts and the butterfly diagram is remarkable, Labonte and Howard (1982) have compared the belts and the latitudinal distribution of the emerging magnetic field. Snodgrass (1985) applied a corrected mathematical procedure and pointed out that the wave do not start from the poles but from lower latitudes. Snodgrass and Dailey (1996) also detected a torsional pattern in the motions of magnetic fields by correlating magnetograms and concluded that this pattern is a result of the meridional outflow. A further velocity field, the vorticity patterns related to the active region outflow has been studied by  Brown and Snodgrass (2003), they conclude that these outflows can also be the sources of the torsional wave. Furthermore, Ulrich (2001) detected wavelike patterns 
 superposed on the torsional belts. The pattern can be detected down to about 0.92 R$_{\odot}$, (Howe et al., 2000, Komm et al. 2001).

The first theoretical attempts considered the Lorentz force (Yoshimura, 1981, Sch\"ussler, 1981) on a large scale. Later small-scale models were put forward by several authors. K\"uker et al. (1996) considered that the magnetic quenching of the Reynolds stresses by the toroidal field may locally modify the differential rotation profile. Spruit (2003) suggested that the cool sunspots generate geostrophic flows resulting in the torsional oscillation. In the model of Petrovay and Forg\'acs-Dajka (2002) the sunspots modify the turbulent viscosity in the convective zone which leads to the modulation of the differential rotation. 

The present work focuses on the spatial correlation of sunspots and the torsional belts. Earlier works pointed out some spatial connections but only using magnetograms (LaBonte and Howard, 1982), Zhao and Kosovichev 2004). By using sunspot data, one can study the role of the most intensive magnetic fluxes and perhaps find an answer to the question: how can sunspot regions be able to modify the ambient flow resulting in the observed zonal velocity pattern?

\section{Observational data}

The shape of the butterfly diagram has been compared with the torsional belts. The velocity distribution has been taken from the paper of Ulrich and Boyden (2005). The source of the sunspot data is the most detailed sunspot catalogue, the Debrecen Photoheliographic Data (DPD, Gy\H ori et al, 2010), the considered period covers the interval 1986-2002. The number of sunspot groups was computed in the following way: the numbers of all spots have been added up in 1 degree wide stripes and 3-month periods in such a way that each sunspot group was taken into account at the time when it contained the largest number of spots, in other terms, when it was in the most developed state. 

The direct comparison of two fractal-like distributions would be difficult or almost impossible by simply ovelapping them, therefore a simplifying tool was used. Separating lines were drawn onto the torsional wave pattern between the prograde and retrograde belts and these lines were inserted into the Schwabe diagram of sunspot numbers. The lines help to find correspondences between the two distributions, see Fig.1.

    \begin{figure}[t]
  \begin{center}
   \epsfig{file=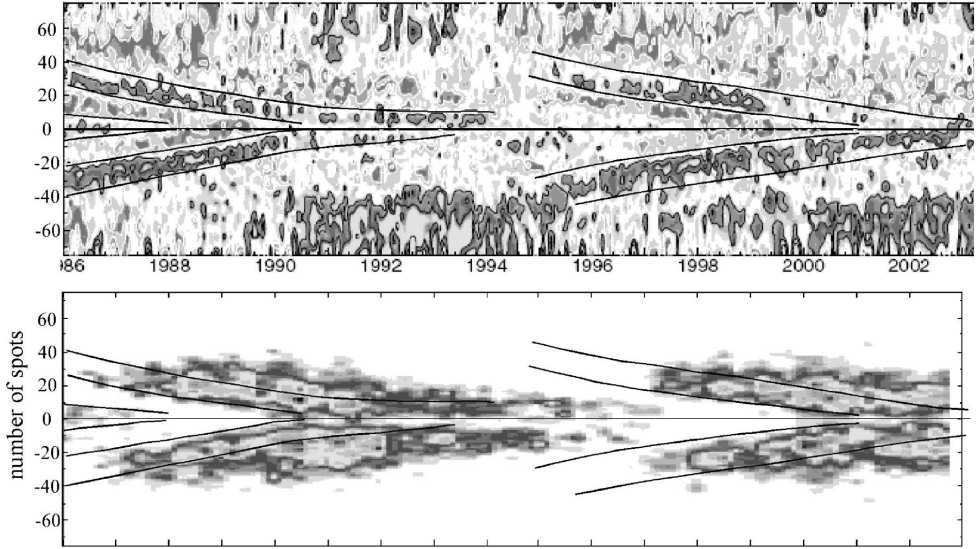,width=12.7cm}
  \end{center}
  \caption{The upper panel shows the azimuthal velocity data of Ulrich and Boyden (2005) along with additionally inserted borderlines of the prograde belts. The second panel shows the distributions of the number of spot groups, the borderlines are inserted.}
\end{figure} 

\section{A possible flow pattern producing torsional belts}

By comparing the two panels of Fig.1. the following properties can be observed. The onset of the torsional wave precedes the appearance of the first spots of the cycle. The speed of approaching the equator is the same for the two phenomena, i.e. the activity belts (the butterfly-diagram) and the torsional belts. The line of weight of the area occupied by sunspots is driving along the poleward borderline of the prograde belt. Perhaps the most interesting property is the definite coincidence of the equatorward borderlines of the prograde belt and the sunspot occurrence. 

The similarity of the two patterns is remarkable but the question of suspected causality connection between spots and torsional waves seems to be inappropriate. We suggest a different approach: the cause of the torsional waves may be the toroidal magnetic flux rope modifying the ambient emerging flow pattern, see Fig.2. If one assumes that the toroidal flux is not restricted to the bottom of the convective zone but some clusters may be present higher up then these clusters would brake the upward motion but outside of the azimuthal clusters (i.e. at their poleward and equatorward sides) these streams are undisturbed. These undisturbed streams would flow around the azimuthal clusters and converge above them. This convergig motion from the polar/equatorial side results in eastward/westward turn respectively due to the Coriolis force.

   \begin{figure}
  \begin{center}
   \epsfig{file=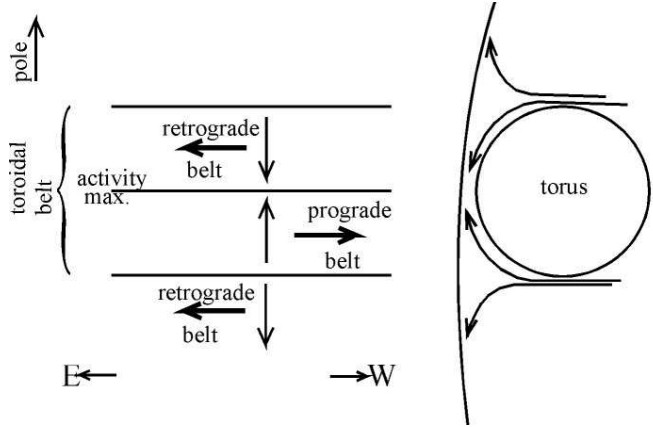,width=9cm}
  \end{center}
  \caption{Schematic representation of the proposed mechanism in the northern hemisphere. Left: view from outside, right: view in the meridional plane from the west.}
\end{figure}

\section{Discussion}

The dilemma of causality is a problem from the beginnings of the theoretical treatment of torsional waves. The first publication of Howard and LaBonte (1980) raised the idea that this velocity field could be the reason of the activity cycle. It turned out soon that this velocity field is too weak to have any impact on the magnetic fields. The reverse cases based on sunspots was not less problematic because the torsional belts are observable even in the absence of any sunspots prior to the beginning of the activity cycle. The present scenario seems to avoid these problems. It only needs the presence of the toroidal flux ropes which is a plausible assumption even before the appearance of the first spots.

\section*{Acknowledgements} 
The research leading to these results has received funding from the European Community's Seventh Framework Programme (FP7/2007-2013) under grant agreement No. 218816.


\section*{References}
\begin{itemize}
\small
\itemsep -2pt
\itemindent -20pt

\item[] Antia, H. M., \& Basu, S., 2001, {\it \apj}, 559, L67
\item[] Gy\H ori, L., Baranyi, T., Ludm\'any, A. et al., 2010 Debrecen Photoheliographic Data http://fenyi.solarobs.unideb.hu/DPD/index.html
\item[] Howard, R., \& LaBonte, B. J., 2005, {\it \apj}, 239, L33
\item[] Howe, R., Christensen-Dalsgaard, J., Hill, et al, 2000, {\it \apj}, 533, L163
\item[] Komm R.W., Hill F., \& Howe R., 2001, {\it \apj}, 558, 428 
\item[] Kosovichev, A. G., \& Schou, J., 1997, {\it \apj}, 482 L207
\item[] K\"uker, M., R\"udiger, G., \& Pipin, V. V., 2005, {\it \aa}, 312, 615
\item[] LaBonte, B. J. \& Howard, R., 1982, \textit{Solar Phys}, 75, 161 
\item[] Petrovay, K., \& Forg\'acs - Dajka E., 2002, \textit{Solar Phys.}, 205, 39 
\item[] Schou, J., Antia, H. M., Basu, S., et al., 1998, {\it \apj}, 505, 390
\item[] Sch\"ussler, M., 1981, {\it \aa}, 94, L17
\item[] Ulrich R. K. \& Boyden, J. E., 2005, {\it \apj}, 620, L123 
\item[] Yoshimura, H., 1981, {\it \apj}, 247, 1102 
\item[] Zhao, J., \& Kosovichev, A. G., 2004, {\it \apj}, 603, 776 
\item[] 
\item[] 

\end{itemize}

\end{document}